\documentclass[aps,preprint,a4]{revtex4-1}

\usepackage{amsmath}
\usepackage{epsfig}
\usepackage{bm}
\usepackage{multirow}
\usepackage{bm}
\usepackage{xcolor}
\usepackage{amsmath}
\usepackage{amssymb}

\begin{document}
\title{
Exact generator and its high order expansions in the
time-convolutionless generalized master equation: Applications
to the spin-boson model and exictation energy transfer
}

\author{Yanying Liu}
\author{Yaming Yan}
\author{Meng Xu}
\author{Kai Song}
\author{Qiang Shi}
\email{qshi@iccas.ac.cn}
\affiliation{Beijing National Laboratory for
Molecular Sciences, State Key Laboratory for Structural Chemistry of
Unstable and Stable Species, CAS Research/Education Center for
Excellence in Molecular Sciences, Institute of Chemistry, Chinese Academy
of Sciences, Zhongguancun, Beijing 100190, China, and
University of Chinese Academy of Sciences, Beijing 100049, China}

\begin{abstract}
The time-convolutionless (TCL) quantum master equation provides a
powerful tool to simulate reduced dynamics of a quantum system
coupled to a bath. The key quantity in the TCL master equation is the
so-called kernel or generator, which describes effects of the bath
degrees of freedom. Since the exact TCL generators are usually hard to
calculate analytically, most applications of the TCL generalized
master equation have relied on approximate generators using second
and fourth order perturbative expansions.
By using the hierarchical equation of motion (HEOM) and extended HEOM methods,
we present a new approach to calculate the exact TCL generator and its
high order perturbative expansions. The new approach is applied to the
spin-boson model with different sets of parameters,
to investigate the convergence of the high order expansions of the
TCL generator. We also discuss circumstances where the
exact TCL generator becomes singular for the spin-boson model, and a model
of excitation energy transfer in the Fenna-Matthews-Olson complex.

\end{abstract}
\maketitle

\section{Introduction}

Quantum dynamics plays a significant role in many chemical and
physical process of condensed phases, while the
development of accurate and efficient schemes to simulate
such processes remains an important challenge in theoretical
chemistry.\cite{nitzan06,weiss12,berne98,may11}
Numerically exact approaches\cite{makarov94,makri95a,
makri95b,wang01,wang03,tanimura89,ishizaki09}
are usually limited to some simple models,
while approximate approaches\cite{tully98,tully12,
thoss01,berkelbach12,montoya15} generally have limited range of
applicability. Therefore,
searching for methods that are accurate, efficient, and general
still stays in the forefront of quantum dynamics in condensed phases.
A popular idea to overcome this issue is to split the total system
in two parts: the quantum system which we are most interested in,
and the environment or bath that exchanges energy and particles with
the system. The quantum system is described specifically using the
reduced density operator (RDO) and its interaction with the bath
is included in the equation of motion of the system
RDO.\cite{pollard96,breuer02,yan05,blum13}

For calculations of the dynamics within the reduced system
dynamics framework, the Nakajima-Zwanzig (NZ) (or time-convolution,
TC) master equation, an integro-differential equation with
a super-operator called ``memory kernel" derived from
the full Hilbert space using the projection operator techniques,
%provides a useful starting point for lots of methods
%based on generalized quantum master equations (GQMEs).
is often employed.\cite{nakajima58,zwanzig60,zwanzig64,mori65}
With the NZ generalized master equation, the complexity of
treating the bath effects is reduced to the calculation of the memory
kernel which completely determines
the dynamics of the system.\cite{breuer02} Although formally exact
memory kernels can be written using the projection operators,
they are hard to compute explicitly except for several simple
models. For example, no analytical exact kernels are available for the
commonly used spin-boson\cite{leggett87,weiss12} and Anderson
impurity\cite{mahan13,gull11} models. Therefore a large number of works
using NZ master equation have been based on perturbation expansions
with respect to the strength of the system-bath coupling or a small coupling
parameter in the system Hamiltonian.\cite{royer03,reichman97,aihara90}
It is noted that, numerical methods to calculate exact NZ memory 
kernel,\cite{shi03,wilner14,cerrillo14} are proposed in many recent studies.
In a recent work, we have also investigated systematically
the convergence of high order expansions of the NZ memory kernel.\cite{xu18}

Instead of using an integro-differential equation to describe the bath effects,
an alternative form of the generalized master equation is
the Hashitsume-Shibata-Takahashi or time-convolutionless (TCL)
approach.\cite{tokuyama76,hashitsumae77,shibata77} Also being
formally exact, this
approach describes the evolution of the system RDO using a first order
differential equation with time dependent
coefficients.\cite{hashitsumae77,shibata77,
chaturvedi79,breuer02} The key quality in the TCL generalized master equation
is a super-operator referred to as the TCL kernel or generator,
which eliminates the
integration over the system history and is local in time.\cite{breuer02}
However, the exact calculation of the TCL generator is
intractable for it relies on performing the inversion of
a super-operator in the
full Hilbert space.\cite{breuer02} Closed expressions
for TCL master equations have been obtained for several
analytically solvable models such as a harmonic oscillator coupled
bilinearly to a harmonic bath,\cite{weiss12,yan05,grabert88,
garraway97} the Jaynes-Cummings model,\cite{murao95,smirne10} and the
resonant energy level model in charge transport.\cite{zhang12}
Although there are recent attempts to obtain the non-perturbative TCL
generators numerically,\cite{nan09,kidon15}
many applications of the TCL generalized master equations
rely on perturbative schemes,
especially the second order perturbation approximations
due to their simplicity.
%These approaches tremendously
%improve the theoretical description of quantum dissipative dynamics
%but have their limitations.
A major problem of the second order TCL master equations is that
their validity is not guaranteed in the case of strong coupling.
The analytical fourth order perturbation approaches,\cite{jang02,breuer02}
however, are usually rather cumbersome and their extensions to
higher orders are difficult. An attempt to formulate
a recursive approach to calculate high orders of the TCL generator
has been proposed recently,\cite{gasbarri18} but has not been applied
to realistic calculations of reduced quantum dynamics.

It should be noted that, when solved exactly, the TCL and NZ master
equations give the same results.\cite{chruscinski10}
While comparing with the NZ form of
the generalized quantum master equation (GQME),
the TCL form of the GQME
usually has wider applications in several problems such as in calculating
spectroscopic signals.\cite{egorova03,schroder06} Especially,
the second order TCL GQME becomes exact for the pure dephasing model
where a two level system with zero interstate coupling is coupled
linearly to a harmonic
bath,\cite{mukamel95,doll08} while the second order NZ GQME is not.
On the computational cost to obtain the memory kernels and generators in
the two forms of GQME, it is found that both kernels/generators usually decay on
similar time scale and thus, on this point, neither formulations is more
advantageous than the other.

In this work, we adopt a population propagator approach
based on recent works\cite{nan09,smirne10,kidon15}, and the hierarchical
equation of motion (HEOM) method\cite{tanimura89,ishizaki05,shi09b}
to calculate the exact TCL generator.
Then, similar to the case of the high order expansions
of the NZ memory kernel,\cite{xu18}
a new approach to calculate the high order expansions of the TCL generator
is proposed based on the previously developed extended HEOM method
to calculate perturbation expansion of open system dynamics.\cite{xu17}
The new approach utilizes the results from the extended HEOM,
but does not involve the cumbersome multi-dimensional integrals.
It is then applied to the spin-boson model to
investigate the convergence of the high order TCL generators.
We also show that, in several examples, the exact TCL generators
may become singular, which may limit the application of the TCL
generalized master equations in certain circumstances.

The remainder of this paper is organized as follows.
In Sec. \ref{tcl} and \ref{sb-heom},
we briefly review the TCL generalized master equation, the spin-boson model
and the HEOM method. Methods to derive the exact
TCL generator and its high order expansions are given
in Sec. \ref{generator}. In Sec. \ref{results},
we apply the proposed method to the spin-boson model and present the numerical
results for the exact TCL generators and their high
order expansions. Examples demonstrating the singularity of the TCL generator
are also presented. Conclusions and discussions are made in Sec. \ref{conclusion}.

\section{Theory}
\label{method}

\subsection{The time-convolutionless master equation}
\label{tcl}
We consider a general total Hamiltonian describing a system
coupled to a bath
\begin{equation}
H = H_S  + H_B + H_{SB} \;\; ,
\label{eq-hamiltonian}
\end{equation}
where $H_S$ is the system Hamiltonian,
$H_B$ is the bath Hamiltonian, and $H_{SB}$ denotes the
system-bath coupling.
Evolution of the total system and bath density matrix is given
by the quantum Liouville equation,
\begin{equation}
\frac{d}{dt}\rho_T(t) = -i[H, \rho_T(t)]
= -i{\cal L} \rho_T(t) \;\; ,
\end{equation}
where ${\rho}_T$ is the density operator of the total system,
${\cal L} \rho_T(t)=[H_T,\rho_T(t)]$, and we set $\hbar =1$
throughout this paper. The reduced system density operator $\rho_S(t)$
is defined as the partial trace over the bath degrees of freedom,
$\rho_S(t) = {\rm Tr}_B {\rho}_T(t)$.

To derive the generalized master equation, we separate the density operator
into relevant and irrelevant parts by means of a projection operator $\mathcal{P}$.
The residual projection operator ${\mathcal Q}$ is defined
as ${\mathcal Q} = 1 - {\mathcal P}$. Both of ${\mathcal P}$ and
${\mathcal Q}$ obey the fundamental requirement of a projection,
${\mathcal P}^2 = {\mathcal P},{\mathcal Q}^2 = {\mathcal Q}$.
Then the TCL form of GQME \cite{hashitsumae77,shibata77,breuer02}
can be written as
\begin{equation}
\label{eq-tclprojector1}
\frac{d}{dt}{\mathcal P} \rho_T(t) =
-i\mathcal{PLP}\rho_T(t)
-i\mathcal{PL}\theta(t)\mathcal{P}\rho_T(t)
+ {\cal I}(t){\mathcal Q}  \rho_T(t_0)\;\; ,
\end{equation}
where
\begin{equation}
{\theta}(t) = {\left[1+i \int_{t_0}^t  ds e^{-i {\mathcal Q} {\cal L}
(t-s)}{\mathcal Q} {\cal L} {\mathcal P}e^{i {\mathcal Q} {\cal L}
(t-s)}\right]}^{-1} \;\; ,
\end{equation}
\begin{equation}
{\cal I}(t) = -i{\mathcal P} {\cal L} \theta (t) e^{-i {\mathcal Q}
{\cal L} (t-t_0)} {\mathcal Q} \;\; ,
\end{equation}
and $\rho_T(t_0)$ is the initial density operator
of the total system.
In this work, we choose a factorized initial condition
$\rho_T(0) = \rho_S(0) \otimes \rho_B(0)$ (we set $t_0$=0),
and assume that ${\mathcal P} \rho_T(0)= \rho_T(0)$,
such that the inhomogeneous term ${\mathcal I}(t)$ vanishes.\cite{nan09}
It is also assumed that $\mathcal{PL}\mathcal{P} = 0$, which holds
for the projection operator used later.
Eq. (\ref{eq-tclprojector1}) thus can be written as
\begin{equation}
\label{eq-tclprojector2}
\frac{d}{dt}{\mathcal P}\rho_T(t)
= -i\mathcal{PL}\theta(t)\mathcal{P}\rho_T(t)\;\; .
\end{equation}

In the following parts of this work, we adopt the
Liouville space notation\cite{sparpaglione88,
sparpaglione88a,chen16}
for the total density operator,
$ \rho_T \equiv \;| \rho_T \rangle \rangle $,
and define the product $\langle \langle A | B
\rangle \rangle \equiv \; {\rm Tr}_S {\rm Tr}_B \{A^\dagger B\}$
where ${\rm Tr}_S$ and ${\rm Tr}_B$ are partial traces over the
system and bath degrees of freedom, respectively.
We denote an eigenstate of the quantum subsystem by
$| j \rangle $,  a Liouville space state of the quantum subsystem
is then given by $| jk \rangle \rangle = | j\rangle \langle k|$.
For simplicity, we use $|j\rangle\rangle$ to denote $|jj\rangle\rangle$, and
$\langle\langle j| = \{|j\rangle\rangle\}^{\dag}$.
The population of the quantum subsystem on state $|j\rangle$
can be written as
$P_{j}(t) ={\rm Tr}_{B}\{|j\rangle\langle j|\rho_{T}(t) \}
=\langle\langle j|\rho_{T}(t)\rangle\rangle$.

We employ the following projection operator
which has been used in many previous studies\cite{chen16,
mavros14,sparpaglione88,zwanzig61}
\begin{equation}
\label{projector}
{\mathcal P} = \sum \limits_j | j \rho_j^B \rangle \rangle
\langle \langle j | \;\;,
\end{equation}
where $| j\rho_j^B\rangle\rangle = | j \rangle \langle j | \otimes \rho_j^B $,
and the locally equilibrated (to be specified later)
bath density operator $\rho_j^B $
is taken to be associated with  the system eigenstate
$|j\rangle $. Thus in the reduced subsystem space,
Eq. ({\ref{eq-tclprojector2}}) can be written as
\begin{equation}
\label{eq-tclpopulation1}
\frac{d}{dt}P (t) = {\cal R}(t)P (t)\;\;,
\end{equation}
where $P(t)$ is a vector whose $j^{th}$ element is the population
on state $| j \rangle$,
and $\mathcal{R}(t)$ is now a matrix whose elements are defined by
${\cal R}_{jk}(t) = -i\langle\langle j|\mathcal{L}\theta(t)
|k\rho_{k}^{B}\rangle\rangle$.
Considering the time evolution of ${\rho}_T(t)$, we have
\begin{equation}
{\mathcal P}{\rho}_T (t)
= {\mathcal P}e^{-i{\cal L} t} {\rho}_T(0)
= {\mathcal P}e^{-i{\cal L} t}
({\mathcal P}+{\mathcal Q}) {\rho}_T(0)\ .
\label{eq-p}
\end{equation}
With the above projector operator $\cal P$ and the relationship
${\cal Q} {\rho}_T(0) = 0$,
and in the reduced subsystem space, Eq. (\ref{eq-p}) can be written as
\begin{equation}
P(t) = {\cal U}_S(t)P(0)\ ,
\label{eq-evolution1}
\end{equation}
where matrix elements of $\mathcal{U}_{S}$,
\begin{equation}
{\cal U}_{S;jk}(t) = \langle\langle j| e^{-i{\cal L} t} |k\rho_k^{B}\rangle\rangle \;\;,
\label{eq-usdefinition}
\end{equation}
involves the propagator of the system only.
Taking time derivative operation on $P(t)$ in Eq. (\ref{eq-evolution1})
and making the inversion of ${\mathcal U}_S(t)$, we get
\begin{equation}
\dot{P}(t) = \dot{{\cal U}}_S(t)P(0) = \dot{{\cal U}}_S(t)
{\cal U}_S^{-1}(t)P(t)\ ,
\label{eq-tclpopulation2}
\end{equation}
provided the inverse does exist.
%This equation has the form of
By comparing with the TCL quantum master equation in
Eq. (\ref{eq-tclpopulation1}),
the TCL generator ${\cal R}(t)$ is given by
\begin{equation}
\label{eq-rt}
{\cal R}(t) =  \dot{{\cal U}}_S(t){\cal U}_S^{-1}(t)\ .
\end{equation}
Thus, in order to obtain the TCL generator, one has to
compute the time derivation of ${\cal U}_S(t)$ and the time inversion of
${\cal U}_S(t)$.
To get the concrete form of $\dot{\cal U}_S(t)$, we take
the derivative of Eq. (\ref{eq-usdefinition}) and obtain:
\begin{equation}
\dot{{\cal U}}_{S;jk}(t)
= -i\langle\langle j| e^{-i{\cal L} t}\mathcal{L} |k\rho_k^{B}\rangle\rangle \;\;.
\label{eq-dus}
\end{equation}
In the above derivations, the condition ${\cal P} {\cal L} {\cal P} = 0 $ have been used.

\subsection{The spin-boson model and HEOM method}
\label{sb-heom}
For simplicity, we restrict ourselves
to the spin-boson model for most part of this paper.
The spin-boson Hamiltonian, in the form of
Eq. (\ref{eq-hamiltonian}), describes a two state system.
The Hamiltonian of the system
\begin{equation}
H_S = {\epsilon} {\sigma}_z + {\Delta} {\sigma}_x \ ,
\end{equation}
is characterized by the energy bias $\epsilon$ and the coupling of
the two states $\Delta$. The Hamiltonian of harmonic bath and
the coupling to the system are
\begin{equation}
H_B = \sum\limits_j \left(\frac{p_j^2}{2} + \frac{1}{2}
{\omega}_j^2 x_j^2 \right) \;\;,
\end{equation}
\begin{equation}
H_{SB} = - \sum\limits_{j} c_jx_j \otimes {\sigma}_z\equiv -F
\otimes {\sigma}_z\ .
\end{equation}
Here, ${\sigma}_z$ and ${\sigma}_x$ are the Pauli matrices.
$x_j$ and $p_j$ are the mass-weighted coordinate and momentum of
the $j^{th}$ bath mode with frequency ${\omega}_j$, respectively.
The coupling coefficient between the system operator ${\sigma}_z$
and the $j^{th}$ bath mode coordinate $x_j$ is written as $c_j$.
$F = \sum_jc_jx_j$ is the collective bath coordinate.
We set ${\beta}=1/(k_BT)$ throughout this paper.

The system-bath interaction is described by the
spectral density, which is defined as:
\cite{chakravarty84,nitzan06}
\begin{equation}
J(\omega) = \frac{\pi}{2} \sum\limits_j \frac{c_j^2}{{\omega}_j}
\delta (\omega -{\omega_j})\ .
\end{equation}
The correlation function $C(t)$ is related to the spectral
density via the fluctuation dissipation theorem:
\cite{weiss12,shi09}
\begin{equation}
C(t>0) =
\frac{1}{Z_B} {\rm Tr}[e^{-{\beta}H_B}F(t)F(0)]
= \frac{1}{\pi} \int_{-\infty}^{\infty} d{\omega}J
({\omega})\frac{e^{-i{\omega}t}}{1-e^{-{\beta}{\omega}}}\ ,
\label{eq-corelation}
\end{equation}
where $Z_B = {\rm Tr} e^{-{\beta}H_B}$ is the partition
function of the uncoupled harmonic bath.
In this work, we will employ the Debye spectral density
\begin{equation}
J(\omega) = \frac{\eta {\omega}_c \omega}{{\omega}^2+{{\omega}_c}^2}\;\;.
\end{equation}
where $\eta$ describes the coupling strength between system and bath,
$\omega_c$ is the cut-off frequency of the bath.
Thus $C(t)$ in Eq. (\ref{eq-corelation}) can be written into
a sum of exponential decaying functions in time:
\begin{equation}
C(t>0) = \sum\limits_k
d_ke^{-{{\omega}}_kt}\,,
\label{eq-corelation1}
\end{equation}
where ${{\omega}}_0 = {{\omega}_c}$ is the longitudinal
relation constant, ${{\omega}}_k = 2k{\pi}/{\beta}$
is the Matsubara frequency and
\begin{equation}
d_0 = \frac{1}{2}{\eta}{{\omega}_c}[\cot(\eta{\omega}_c/2)-i] \;\;,
\end{equation}
\begin{equation}
d_k = \frac {4k\pi\eta{\omega}_c}{{(2k\pi)}^2
-{(\beta{\omega}_c)}^2}\ ,\, {\rm for}\,\,k\geq1\ .
\end{equation}
Then the HEOM can be derived using the path integral technique,
\cite{tanimura89,tanimura90,
ishizaki05,tanimura06,xu07,weiss12,tanimura94,ishizaki08}
or the stochastic Liouville equation approach
\cite{yan04,zhou05,zhou08}:
\begin{eqnarray}
\frac{\partial}{{\partial}t} {\rho}_{\textbf{n}}(t) =
& - & \left(i{\cal L} + \sum\limits_k n_k {{\omega}}_k\right)
{\rho}_{\textbf{n}}(t) - i\left[ {\sigma}_z, \sum\limits_k
{\rho}_{{\textbf{n}}_k^+}(t)\right] \nonumber \\
& - & i \sum\limits_kn_k \left(d_k {\sigma}_z {\rho}_{{\textbf{n}}_
k^-}(t)- d_k^* {\rho}_{{\textbf{n}}_k^-}(t) {\sigma}_z \right)\ .
\label{eq-heom1}
\end{eqnarray}
The subscript \textbf{n} denotes a set of index
$\{n_1,n_2,...,n_k,...\}$, with the integer number $n_k \geq 0$
associated with the $k^{th}$ exponential terms in $C(t)$ of
Eq. (\ref{eq-corelation1}). The subscript ${\textbf{n}}_k^{\pm}$
differs from \textbf{n} only by changing the specified
$n_k$ to $n_k\pm 1$. ${\rho}_{\textbf{0}}$ with
$\textbf{0} = \{0,0,...\}$ corresponds to the reduced system
density operator and the other ${\rho}_{\textbf{n}}s$ are
the auxiliary density operators (ADOs).

\subsection{The exact TCL generator and its high order expansions}
\label{generator}
As shown in Eq. (\ref{eq-rt}), to obtain the exact TCL generator, we first
need to calculate the value of ${\cal U}_S(t)$, defined in
Eq. (\ref{eq-usdefinition}).
The matrix element of ${\cal U}_S(t)$ can be written as
\begin{eqnarray}
{[{\cal U}_S(t)]}_{jk}
& = & \langle \langle j | e^{-i {\cal L}t} | k
\rho_k^B \rangle \rangle \nonumber \\
& = & {\rm Tr}_S{\rm Tr}_B \{| j \rangle \langle j | e^{-i{\cal L}t}
| k \rangle \langle k| \otimes\rho_k^B \} \nonumber \\
& = & [\sigma_k(t)]_{jj} \;\;,
\label{eq-us2}
\end{eqnarray}
where $\sigma_k(t) = {\rm Tr}_B \{ e^{-i{\cal L}t} | k
\rangle \langle k | \otimes \rho_k^B \}$ and the subscript $jj$
denotes the $j^{th}$ diagonal matrix element of the system
reduced density operator. It is noted that the initial state of bath
to compute $[\sigma_k(t)]_{jj} $ is the relaxed equilibrium associated
with the $k^{th}$ state of system.
 The following projection operator for the
spin-boson model is employed,
\begin{equation}\label{Eq:sb-projction}
   \mathcal{P} = |1\rho_1^{B}\rangle\rangle\langle\langle 1| +
                  |2\rho_2^{B}\rangle\rangle\langle\langle 2| \;\;,
\end{equation}
where $\rho_{j}^{B}$ is defined as the locally equilibrated bath density
operator associated with state $|j\rangle$, and $|1\rangle$ denotes donor
state, $|2\rangle $ denotes acceptor state.
\begin{equation}
   {\rho}_j^B = \frac{e^{-{\beta}H^{(j)}}}{{\rm Tr}_B\{e^{-{\beta}H^{(j)}}\}} \;\;,
\end{equation}
with $H^{(j)} = \pm({\epsilon} + \sum\limits_{\alpha}c_{\alpha}x_{\alpha})
+ H_B$
$(+$ for state $|1\rangle$, $-$ for state $|2\rangle$, respectively $)$.
Therefore, within the spin-boson model,
we can calculate ${\cal U}_S(t)$ using the HEOM approach
presented
in the above subsection. By taking time inversion
of ${\cal U}_S(t)$, we can get ${\cal U}^{-1}_S(t)$.

To calculate the derivative of ${\cal U}_S(t)$, we define
\begin{equation}
H_S = H_0 + H_1
\end{equation}
where $H_0 = \epsilon {\sigma}_z$ and $H_1 = \Delta {\sigma}_x$.
The relation ${\mathcal Q}{\cal L}{\mathcal P} = {\cal L}_1 {\mathcal P}$
holds for the above defined projection operator in
Eq. (\ref{Eq:sb-projction}).
$\dot{\cal U}_{S;jk}(t)$ defined in Eq. (\ref{eq-dus}) can then be
calculated using ${\cal L}_1 | k\rangle\rangle =
{(-1)}^ki{\Delta}{\sigma}_y$,
\begin{eqnarray}
{[\dot{\cal U}_S(t)]}_{jk} = {(-1)}^{j+k+1}{\Delta}{[{\sigma}_z{\sigma}_
{y,k}(t)]}_{jj} \;\;,
\label{eq-us3}
\end{eqnarray}
where ${\sigma}_{y,t}(t) = {\rm Tr}_B\{e^{-i{\cal L}t}{\sigma}
_y{\rho}_k^B\}$.
Similar to ${\cal U}_S(t),\,\dot{\cal U}_S(t)$ can also be
computed using the HEOM method.
Therefore, by calculating $\dot{\cal U}_S(t)$ and ${\cal U}^{-1}_S(t)$, it is easy to get
the value of the exact TCL generator with Eq. (\ref{eq-rt}).

In the rest part of this section, we will deduce the high order
expansions of ${\cal R}(t)$. By using the projection operator in
Eq. (\ref{projector}), the coupling between the two states $\Delta$
is used as the expansion parameter.
We first write the perturbative expansions of ${\cal R}(t)$,
${\cal U}_S(t)$, and $\dot{\cal U}_S(t)$, with respect
to the inter-state coupling ${\Delta}$:
\begin{eqnarray}
{\cal R}(t) = \sum\limits_{N=2}^{\infty}{\Delta}^N{\cal R}^{(N)}(t) \;\;,\\
\label{eq-us5}
{\cal U}_S(t) = {\bf I}+\sum\limits_{N=2}^{\infty}{\Delta}^N
{\cal U}_S^{(N)}(t) \;\;, \\
\label{eq-us4}
\dot{\cal U}_S(t) = \sum\limits_{N=2}^{\infty}{\Delta}^N
\dot{\cal U}_S^{(N)}(t) \;\;.
\end{eqnarray}
By exploiting the identity ${(1+x)}^{-1} = \sum_{n=0}^{\infty}
{(-1)}^nx^n$ (under the assumption that $ | x | < 1$),
we can invert Eq. (\ref{eq-us4}) as
\begin{equation}
{\cal U}_s^{-1}(t) = \sum\limits_{n=0}^{\infty}{(-1)}^n
{\left[\sum\limits_{N=2}^{\infty}{\Delta}^N{\cal U}_S^{(N)}(t) \right]}^n\ .
\end{equation}
Thus we can obtain from Eq. (\ref{eq-rt}),
\begin{equation}
\sum\limits_{N=2}^{\infty}{\Delta}^N{\cal R}^{(N)}(t) =
\left\{\sum\limits_{N=2}^{\infty}{\Delta}^N\dot{\cal U}_
S^{(N)}(t)\right\}
\left\{\sum\limits_{n=0}^{\infty}{(-1)}^n{\left[\sum\limits_
{N=2}^{\infty}{\Delta}^N{\cal U}_S^{(N)}(t) \right]}^n\right\}
 \;\;.
\end{equation}
By expanding this equation and comparing coefficients for each order
of $\Delta$ results in an infinite series of equations:
\begin{eqnarray}
{\cal R}^{(2)}(t)& = & \dot{\cal U}_S^{(2)}(t) \;\;, \\
{\cal R}^{(4)}(t)& = & \dot{\cal U}_S^{(4)}(t) -
\dot{\cal U}_S^{(2)}(t) {\cal U}_S^{(2)}(t)  \;\; ,\\
{\cal R}^{(6)}(t)& = & \dot{\cal U}_S^{(6)}(t) -
\dot{\cal U}_S^{(4)}(t) {\cal U}_S^{(2)}(t) +
\dot{\cal U}_S^{(2)}(t) {\cal U}_S^{(2)}(t) {\cal U}_S^{(2)}(t)
- \dot{\cal U}_S^{(2)}(t) \ {\cal U}_S^{(4)}(t) \;\;, \\
\cdots . \nonumber
\end{eqnarray}
The odd order terms in the expansion of the exact generator ${\cal R}(t)$
are all zero. With the above equations, the following
relations can be obtained for the even order terms:
\begin{equation}
\label{eq-recursion}
{\cal R}^{(2n)}(t) = \dot{\cal U}_S^{(2n)}(t)- \sum\limits_{m=1}^{n-1}
[{\cal R}^{(2m)}(t) {\cal U}_S^{[2(n-m)]}(t)]\ .
\end{equation}

In order to obtain the expansions in Eqs. (\ref{eq-us5}) and (\ref{eq-us4}),
the extended HEOM method\cite{xu17} is employed.
More specifically, the perturbation expansion of ${\cal U}_S(t)$
can be calculated directly using the method presented in Ref.\cite{xu17}
with proper initial conditions. Expansions of $\dot{\cal U}_S(t)$
with respect to $\Delta$ can be calculated by expanding
${\sigma}_{y,k}(t)$ defined in Eq. (\ref{eq-us3}) into Taylor series:
\begin{equation}
{\sigma}_{y,k}(t) = \sum\limits_{N=0}^{\infty} \frac{1}{N!}
{\sigma}_{y,k}^{(N)}(t){\Delta}^N \;\;.
\label{eq-sigma}
\end{equation}
The ${\sigma}_{y,k}^{(N)}(t)$ terms can be obtained using
the extended HEOM method\cite{xu17} with the initial condition
${\rho}_T = {\sigma}_y \otimes {\rho}_k^B$ :
\begin{eqnarray}
\frac{\partial}{{\partial}t}{\rho}_{\textbf{n}}^{(N)}(t) =
&-&i{\cal L}_0{\rho}_{\textbf{n}}^{(N)}(t)-iN[{\sigma}_x,{\rho}_
{\textbf{n}}^{(N-1)}(t)]-\sum\limits_k(n_k{{\omega}}_k)
{\rho}_{\textbf{n}}^{(N)}(t)\nonumber \\
&-&i\left[{\sigma}_z,\sum\limits_k{\rho}_{{\textbf{n}}_k^+}^{(N)}(t)\right]
-i\sum\limits_kn_k\left[c_k{\sigma}_z{\rho}_{{\textbf{n}}_k^-}^
{(N)}(t)- d_k^*{\rho}_{{\textbf{n}}_k^-}^{(N)}(t){\sigma}_z\right] \;\;,
\label{eq-heom2}
\end{eqnarray}
where ${\cal L}_0 {\rho}=[{\epsilon}{\sigma}_z,{\rho}]$.

\section{Results}
\label{results}
For all numerical calculations in this section, the initial
state of the total system is assumed to be equilibrated on
state $| 1\rangle$, i.e., ${\rho}_T(0) = | 1 \rangle
\langle 1| \otimes {\rho}_1^B$. Thus the inhomogeneous term
in Eq. (\ref{eq-tclprojector1}) vanishes. The HEOM approach is used to
calculate ${\cal U}_S(t)$ with Eq. (\ref{eq-us2}) and the
benchmark numerical exact population dynamics. Population
dynamics of TCL quantum master equation is calculated from
the differential equation in Eq. (\ref{eq-tclpopulation1}), which is solved
by the fourth order Runge-Kutta method.
The high order expansion terms of the TCL generator
${\cal R}(t)$ are computed by the extended HEOM approach with
Eqs. (\ref{eq-us3}), (\ref{eq-heom2}), and (\ref{eq-recursion}).

There are four numerical examples in this section. The first three
are based on the spin-boson model with different parameters,
including the system-bath coupling strength $\eta$, the cut-off
frequency ${\omega}_c$, the intra-state coupling $\Delta$, the
inverse temperature $\beta$, and the energetic bias $\epsilon$,
while the last one is for the excitation energy transfer in the
Fenna-Matthews-Olson (FMO) complex. The first two
examples describe circumstances  where the high order expansions
of the TCL generator converge and diverge, respectively.
The possible singularity of the exact generator is shown in the third
example for the spin-boson model and the fourth example for the FMO complex.

\subsection{Convergence of high order expansions of the generator}

\begin{figure}[htbp]
\centering
\includegraphics[width=15cm]{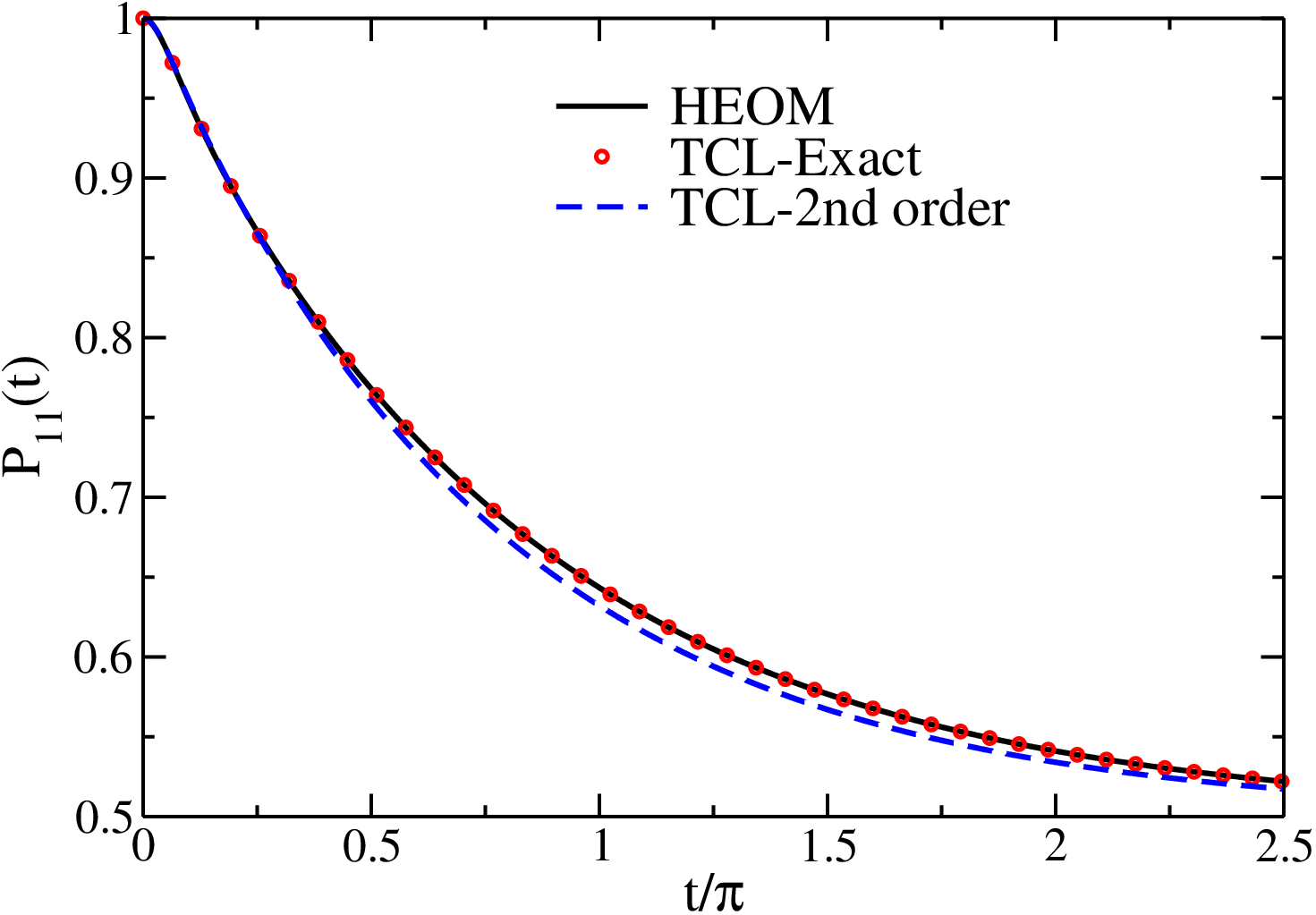}
\vspace{1em}
\caption{The time dependent population on state
$|1\rangle$ in the spin-boson model, obtained from the HEOM method,
the exact TCL generalized master equation, and the TCL equation
with the second order approximate generator,
respectively. The parameters are
${\beta} = 0.5, {\omega}_c = 5, {\eta} = 5, {\Delta} = 1$,
and ${\epsilon} = 0$.}
\label{fig1}
%\vspace{35em}
\end{figure}

In this subsection, we investigate the convergence of the high
order perturbative expansions of the TCL generator presented in
the above Sec. II.C.
In the first example, Fig. \ref{fig1} shows the time evolution
of population calculated by the HEOM method as well as the TCL
generalized master equation with the exact generator.
The parameters in this example are ${\beta} = 0.5,\,
{\omega}_c = 5,\,{\eta} = 5,\, {\Delta} = 1$, and ${\epsilon} = 0$.
It is noted that the population dynamics resulting from the
two different approaches are exactly same, thus validating the
method to obtain the exact generator.
Population dynamics using the second order TCL generalized master
equation is also shown for comparison. In this example, results
from the second order TCL approximation also agree
well the exact ones.

\begin{figure}[htbp]
\centering
\includegraphics[width=15cm]{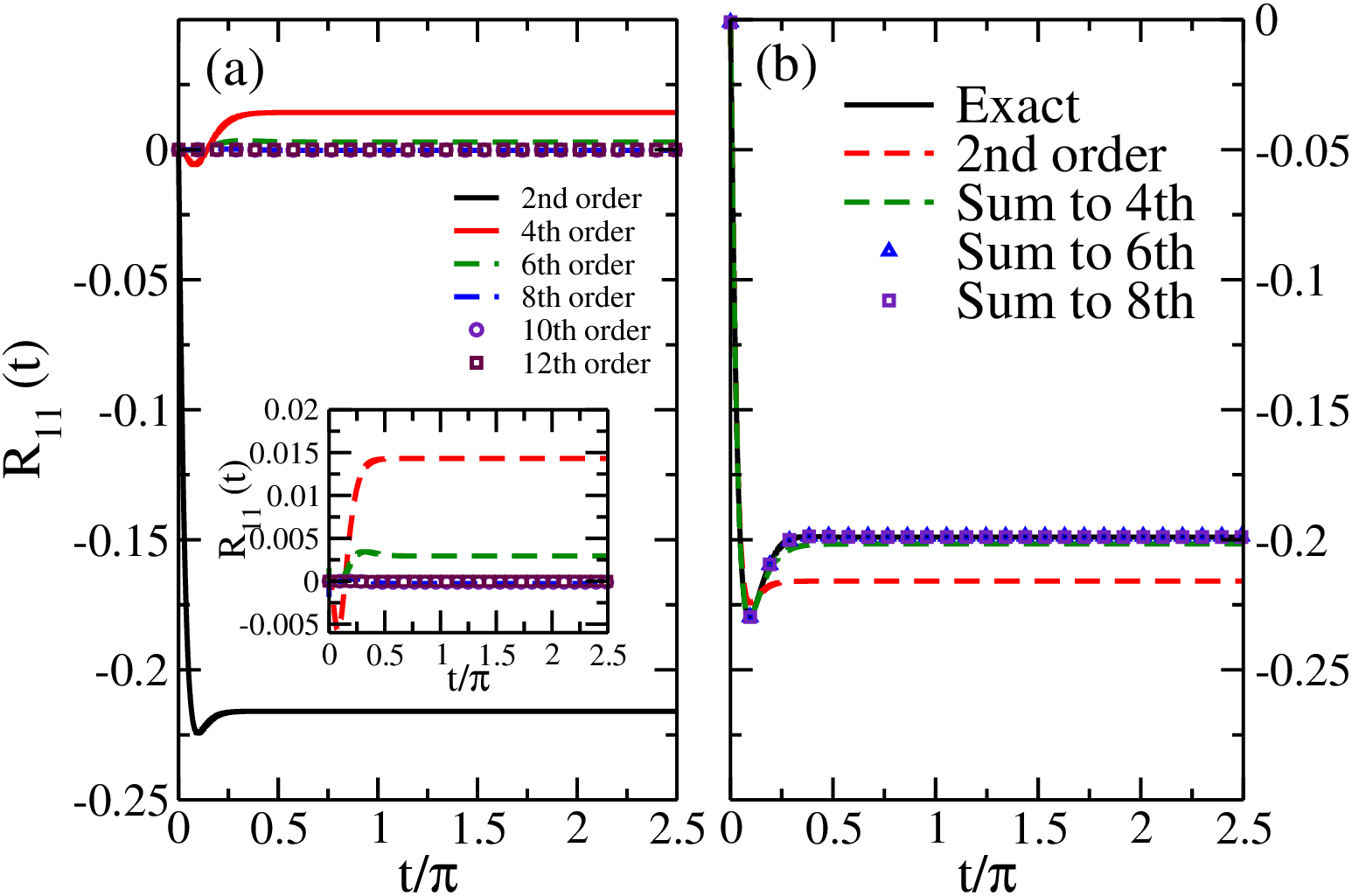}
\vspace{1em}
\caption{Panel (a) shows the second and high order
expansion terms of the TCL generator.
Panel (b) shows the corresponding summations to certain orders.
The parameters are same as those in Fig. \ref{fig1}.}
\label{fig2}
%\vspace{35em}
\end{figure}

The high order expansions up to the 12th order of the TCL
generator ${\cal R}(t)$ are shown in Fig. \ref{fig2}(a),
and the fine details are shown in the inset.
The corresponding summations of $n^{th}$ order generators
are given in Fig. \ref{fig2}(b).
Fig. \ref{fig2}(a) shows that the amplitudes of high order terms
decrease quickly when the perturbation order increases, making the
expansion of the generator converge easily, as shown in
Fig. \ref{fig2}(b).

\begin{figure}[htbp]
\centering
\includegraphics[width=15cm]{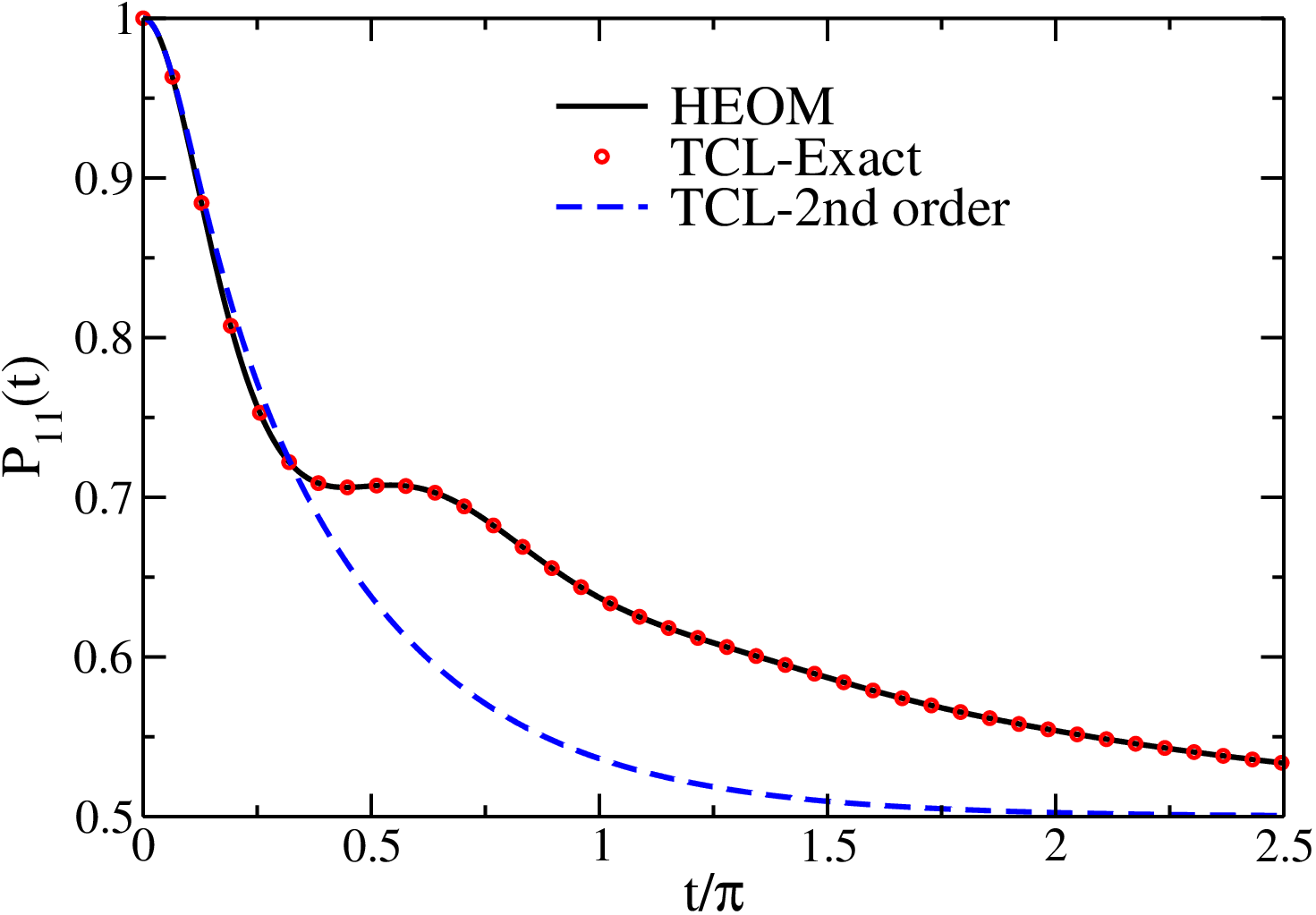}
\vspace{1em}
\caption{Same as Fig. 1, for the parameters:
${\beta} = 1, {\omega}_c = 1, {\eta} = 2, {\Delta} = 1$,
and ${\epsilon} = 0$.}
\label{fig3}
%\vspace\overleftarrow{{35em}}
\end{figure}

In the second example, the following parameters are
used: ${\beta} = 1, {\omega}_c = 1, {\eta} = 2, {\Delta} = 1$,
and ${\epsilon} = 0$, which corresponds to the case of weak
system-bath coupling and a slow bath.
Fig. \ref{fig3} shows the exact population dynamics
using the HEOM method, and the TCL generalized master equations
with the exact and second order generators.
It can be seen that, after a short time, results from
the second order TCL generalized master equation quickly
deviates from the exact ones.

\begin{figure}[htbp]
\centering
\includegraphics[width=15cm]{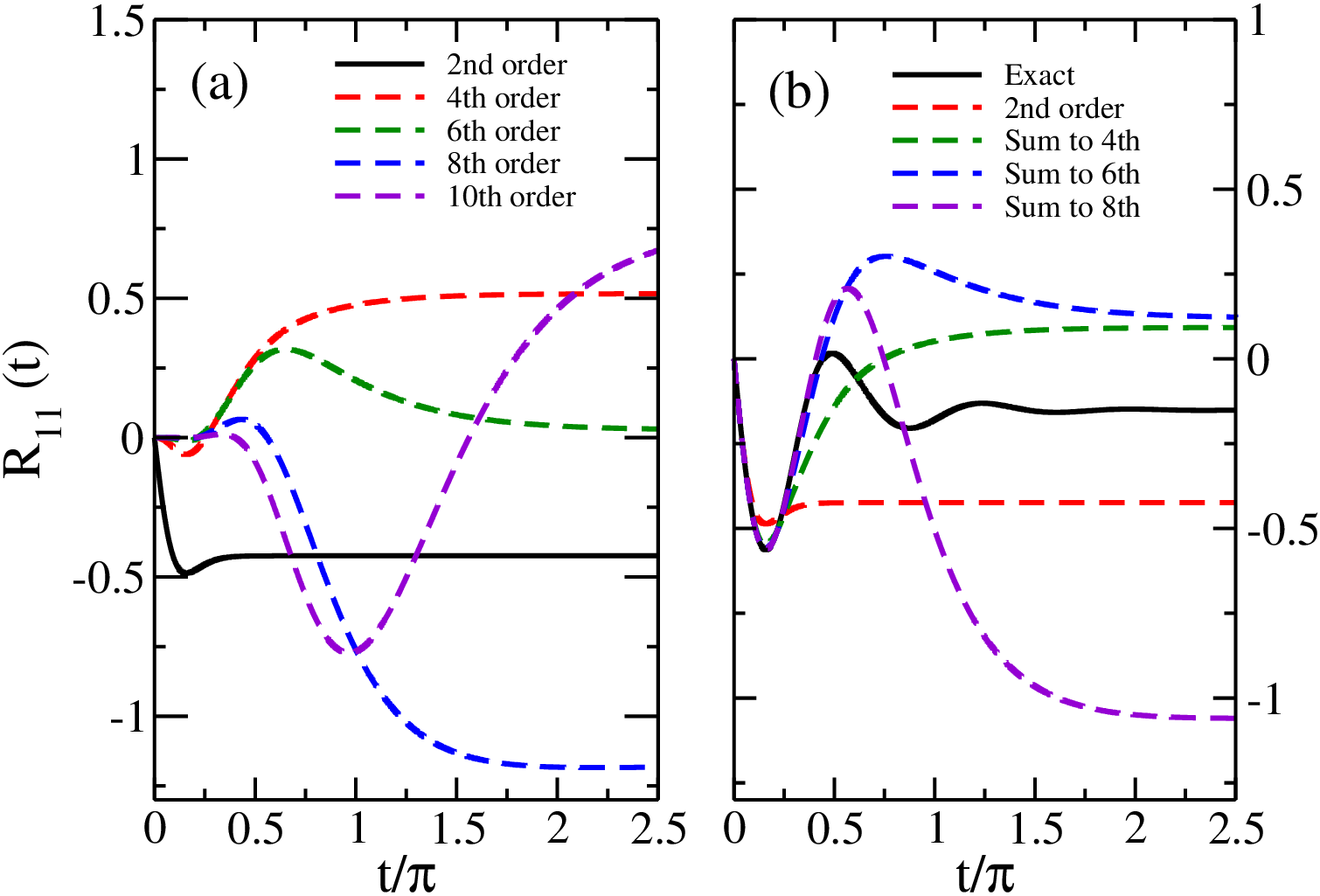}
\vspace{1em}
\caption{Same as Fig. 2, with the parameters same as those in
Fig. \ref{fig3}.}
\label{fig4}
%\vspace{35em}
\end{figure}

The high order expansions of the TCL generator for the second example
is shown in Fig. \ref{fig4}(a). In contrast to the case shown
in Fig. \ref{fig2}(a),
the amplitude of the high order terms
does not decrease as the perturbation order increases.
Thus it is expected that the perturbation expansion at finite order will
soon breakdown when the propagation time increases so that the generator
expansion becomes hard to converge. This is exactly the case
as shown in Fig. \ref{fig4}(b).

By comparing the high order expansions shown in
Fig.\ref{fig2} and \ref{fig4}, a good indication
of the possible convergence is to compare the relative
amplitudes of the second order and fourth order perturbation
terms $\mathcal{R}^{(2)}$ and $\mathcal{R}^{(4)}$.
This has been discussed previously for the NZ form of the
generalized master equations or polaron-transformed 
GQMEs.\cite{jang09,mccutcheon10,chang13,chen16,xu18}
The general trend is also applicable for the TCL generators.

\begin{figure}[htbp]
\centering
\includegraphics[width=15cm]{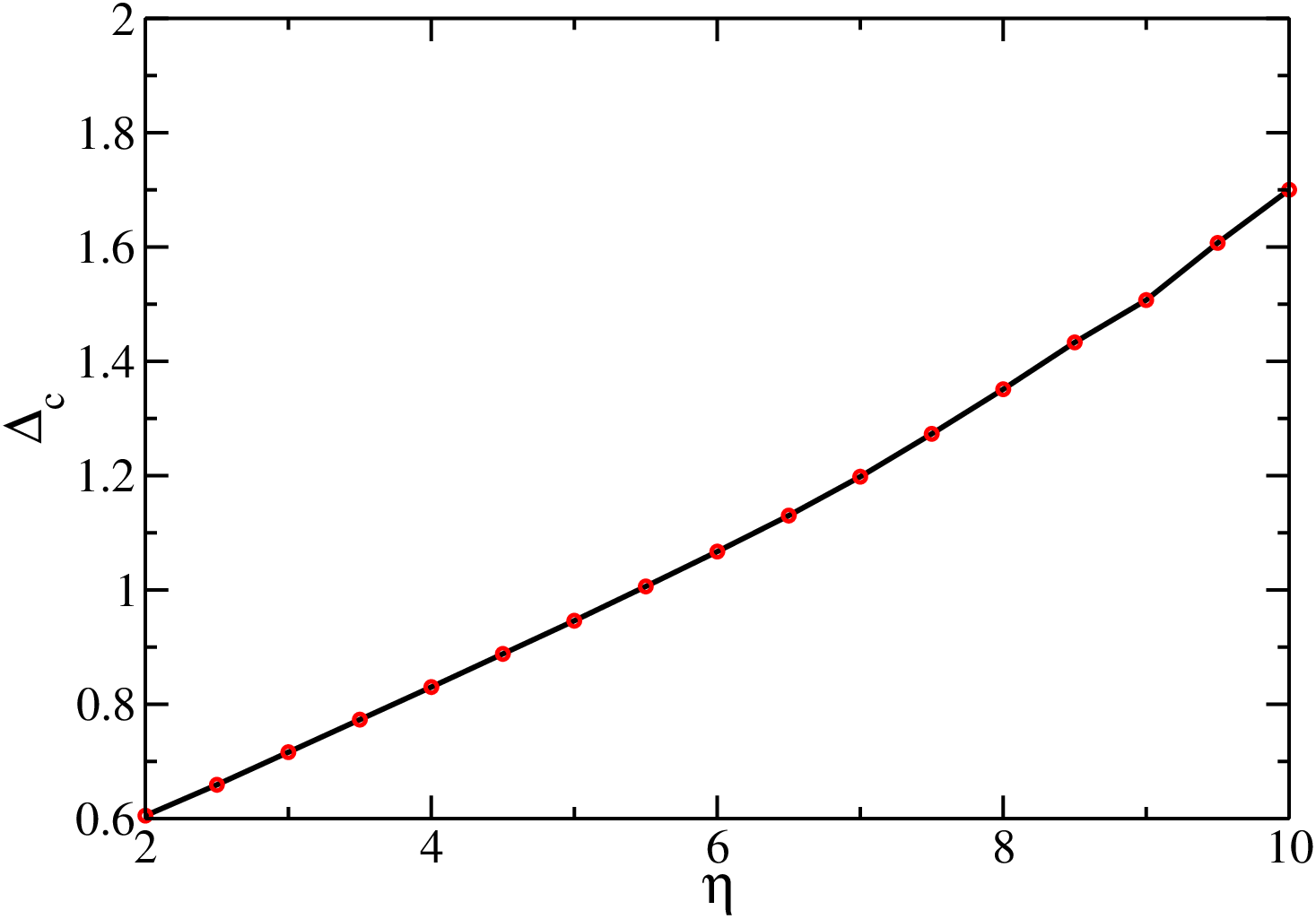}
\vspace{1em}
\caption{The critical interstate coupling constant $\Delta_c$ for converged
expansion of the TCL generator, as a function of the system-bath coupling
strength $\eta$.
The other parameters are
${\beta} = 0.5, {\omega}_c = 5$, and ${\epsilon} = 0$.}
\label{fig5}
%\vspace{35em}
\end{figure}

To explore the convergence of the high order TCL generators
in different parameter regimes, we also calculate the critical values
of interstate coupling ${\Delta}_c$ as a function of the system-bath coupling strength $\eta$,
where the high order expansions converge when $\Delta < \Delta_c$.
%Fig. \ref{fig5} shows
%that the critical value of interstate coupling increase with
%the system-bath coupling.
The convergence criterion is set to
\begin{equation}
|S_{11}^{(n)}(t/\pi=2.5) - S_{11}^{(10)}(t/\pi=2.5)| \leqslant 0.001 \;\; ,
\end{equation}
for the perturbation order $n$ up to 28, where $S_{11}^{(n)}(t/{\pi})$
represents the first matrix element of the summation of the $n^{th}$
order generators at time $t$.
The other parameters we used are ${\beta} = 0.5, {\omega}_c = 5$,
and ${\epsilon} = 0$. The results are shown in Fig. \ref{fig5}.
It can be seen that, with the increase of the
system-bath coupling strength $\eta$, $\Delta_c$ also increases.

\subsection{Singularity of the exact generator ${\cal R}(t)$}

In contrast to the memory kernel in the NZ generalized master equation, the
formally exact generator of the TCL generalized master equation in
Eq. (\ref{eq-tclprojector1}) involves the inverse of a Liouville space
superoperator. And as discussed in Sec. \ref{method}, we assume that
the inverse of ${\cal U}_S(t)$ does exist but there are
some circumstances in which ${\cal U}_S(t)$ is not a full rank matrix,
resulting in the singularity points of the generator at certain time $t$.

\begin{figure}[htbp]
\centering
\includegraphics[width=15cm]{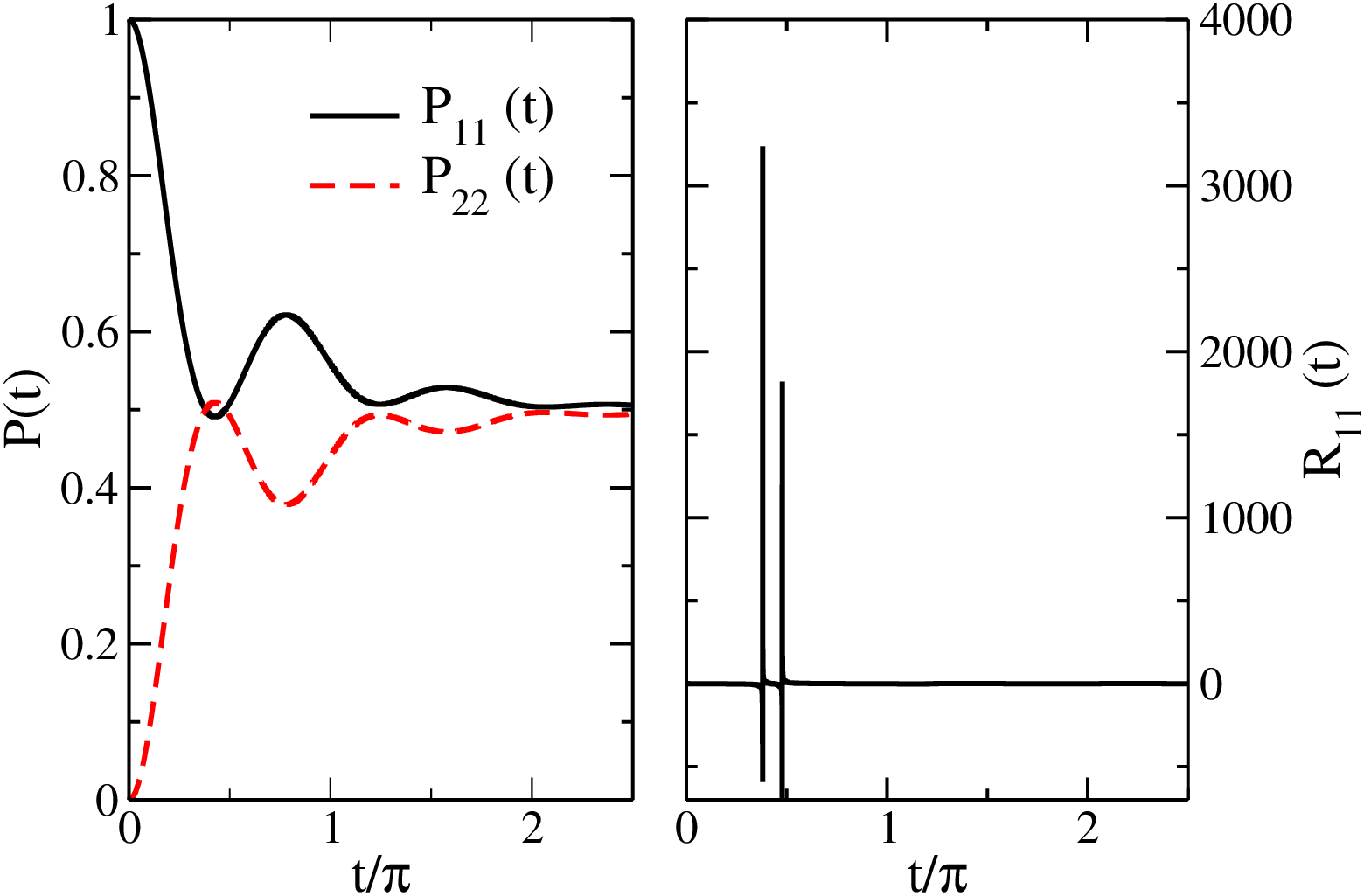}
\vspace{1em}
\caption{Panel (a) shows the time dependent population on state
$|1\rangle$ and $|2\rangle$ in the spin-boson model.
Panel (b) shows the corresponding numerical result of
the exact generator. The parameters are: ${\beta} = 1,
{\omega}_c = 1, {\eta} = 1, {\Delta} = 1$,
and ${\epsilon} = 0$.}
\label{fig6}
%\vspace{35em}
\end{figure}

In this subsection, we show such an example by setting
$ {\omega}_c = 1, {\eta} = 1$, and the other parameters are same
as those in the second example. We can see from Fig. \ref{fig6}(b)
that with this set of parameters,
${\cal R}(t)$ has two singularities. Comparing Fig.\ref{fig6} (a)
with (b), it is noted that when
$P_{11}$ equals to $P_{22}$, the singularities appear.

This result can be understood using the following theoretical analyses.
For the symmetric spin-boson model, the ${\cal U}_S(t)$ and
${\cal R}(t)$ satisfies the following relations:
\begin{equation}
{\cal U}_S(t) =
\left(
  \begin{array}{cc}
    a(t) & 1-a(t)\\
    1-a(t) & a(t)\\
  \end{array}
\right),
\;\;\;\;{\cal R}(t) =
\left(
  \begin{array}{cc}
    b(t) & -b(t)\\
    -b(t) & b(t)\\
  \end{array}
\right).
\label{eq-singularity}
\end{equation}
When $a(t) = 1/2$, the inverse of ${\cal U}_S(t)$ does not exist.
However, in this case, by using the above Eq. (\ref{eq-singularity}),
the TCL generalized master equation
in Eq. (\ref{eq-tclpopulation1}) predicts that $\dot P_i(t) = 0$,
which can not hold except in the equilibrium state.
So for the symmetric spin-boson model, whenever the population curves
cross, there will be singularity in the exact generator.

\begin{figure}[htbp]
\centering
\includegraphics[width=15cm]{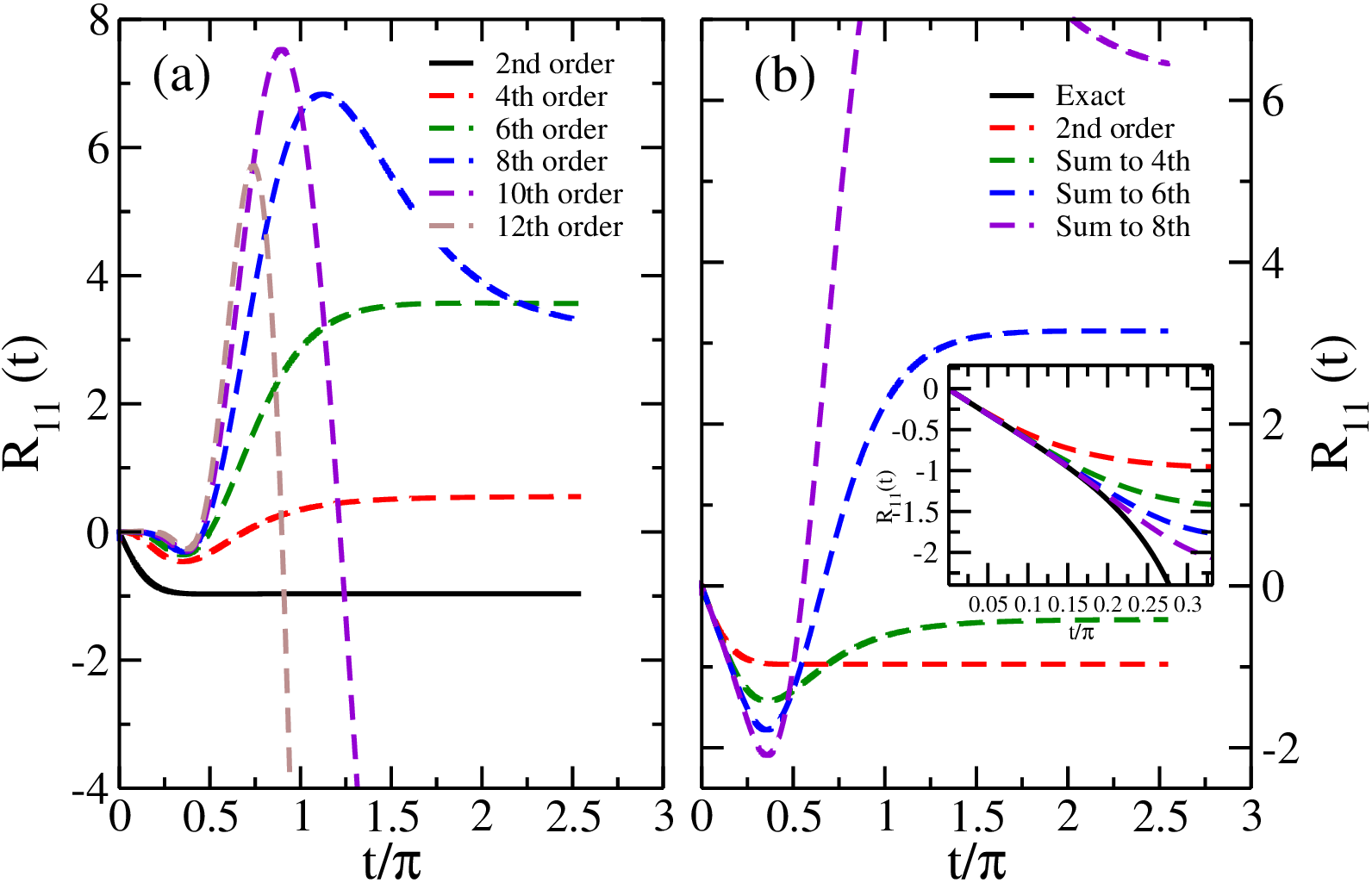}
\vspace{1em}
\caption{Panel (a) shows the second and high order
(4th to 12th) expansion terms of the TCL generator.
Panel (b) shows the corresponding summation terms.
The parameters are same as those in Fig. \ref{fig6}.}
\label{fig7}
%\vspace{35em}
\end{figure}

Because of the existence of the singularities, it is reasonable
to predict the divergence of the high order expansion of the
TCL generator in this example. The results shown in
Fig. \ref{fig7}(a) and (b) verify
this prediction. It can be seen from Fig. \ref{fig7}(a) that
amplitudes of high order terms increase rapidly with the
perturbation order. And the results in Fig. \ref{fig7}(b)
show that the summations of perturbative terms diverge from
the exact one after only a very short time.

\begin{figure}[htbp]
\centering
\includegraphics[width=15cm]{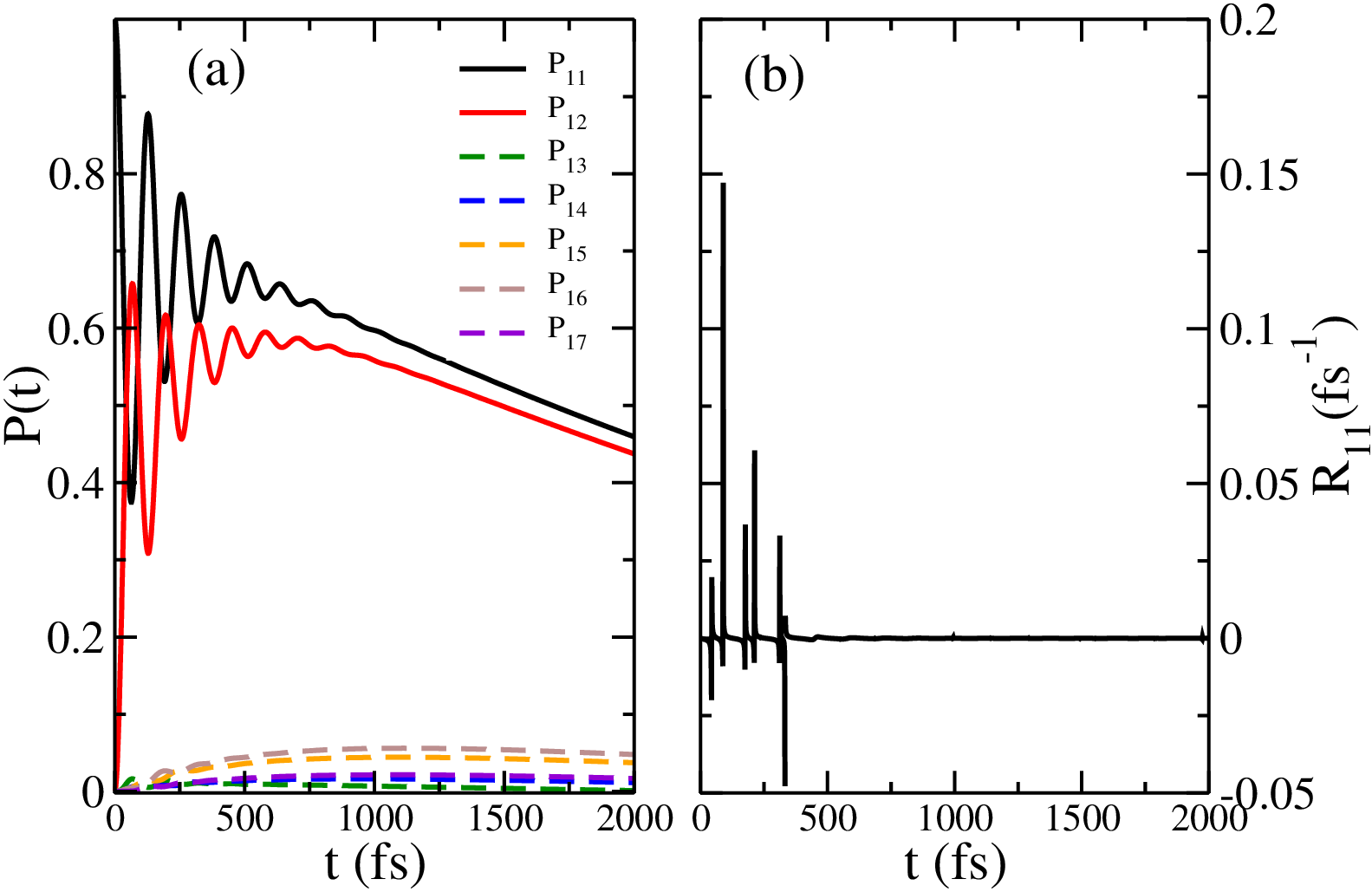}
\vspace{1em}
\caption{Panel (a) shows the time dependent population on sites
1 to 7 of the FMO complex. The initial state is equalibrated on site 1.
Panel (b) shows the corresponding numerical result of
the exact generator. }
\label{fig8}
%\vspace{35em}
\end{figure}

It is interesting to investigate whether similar problem of
singular generators also exists in more general model systems beyond
the simple symmetric spin-boson model.
We apply the method mentioned in Sec. \ref{generator} to the
problem of excitation energy transfer in the FMO complex.
The model Hamiltonian for this problem can be
found in previous works,\cite{cho05} and the parameters used in this
simulation are obtained from Refs.\cite{cho05,vulto98}. Fig. \ref{fig8}(a)
shows the time evolution of population when the initial state is
assumed to be equilibrated on site 1 of the FMO complex. (It is noted
that most simulations in the literature use an unrelaxed initial state
for the bath degrees of freedom) And Fig. \ref{fig8}(b) shows the
corresponding exact generator $\mathcal{R}_{11}$.
It is found that there are also several singularity points in the TCL
generator. So that although the exact TCL generalized master equation
is very appealing in the simulation of open system dynamics, the issue of
singularity in the exact TCL generator may exist in many problems.
In contrast, the memory kernel in the NZ generalized master equation
is usually well behaved.
%the method discussed in Sec. \ref{generator}
%to compute the exact generator failed in FMO complex, due to the
%invalidity of the assumption that the inverse of ${\cal U}_S(t)$ exists.

It should also be noted here that, the above behaviors of singular TCL generator
are investigated using a generalized master equation of populations, using the
projection operator defined in Eq. (\ref{projector})
The condition of singular generator will also be different
if other types of projection operators are employed.

\section{Conclusions and discussions}
\label{conclusion}
In this work, we have combined the population propagator formalism
and the HEOM method to obtain the exact TCL generator,
and proposed a new approach to calculate the high order expansions of
the generator with the recently developed extended HEOM method to
obtain high order expansions of the open system dynamics.
A recursive relation is derived with which one can obtain the high
order expansions of the TCL generator without the need
to calculate the cumbersome and time-consuming multi-dimensional
integrals.

By using the spin-boson model as an example, we have investigated the
convergence of the high order expansion of the TCL generator. It
is shown that the convergence of the high order generators depends
on the system parameter, and the ratio between the fourth and second
order terms can serve as a good indicator of the convergence.
A potential problem of the exact TCL generator is that, it may become
singular in certain parameter regimes, as has been demonstrated
in the analytically solvable Jaynes-Cummings model\cite{smirne10}
and the resonant energy model\cite{zhang12}
for charge transport. We have also shown that the problem of singular
exact generator also exists in the case of spin-boson model,
and a commonly used model for excitation energy transfer in the FMO
complex. Of course, the condition for singular generator also depends
on the choice of projection operator, as well as the parameters. The current
study is based on the TCL generalized master equation using system populations,
and this topic may be further explored for other different forms of TCL GQME.

\acknowledgements
This work is supported by NSFC (Grant No. 21673246),
and the Strategic Priority Research Program of the Chinese Academy
of Sciences (Grant No. XDB12020300).

\pagebreak

\pagebreak

\pagebreak

\pagebreak

\pagebreak

\pagebreak

\pagebreak

\pagebreak

\end{document}